\documentclass[a4paper,UKenglish]{lipics}
\usepackage{microtype}%if unwanted, comment out or use option "draft"

\usepackage{amsmath, amssymb}
\usepackage{color}
\usepackage{enumerate}
\usepackage{xspace}

\usepackage{graphicx}
\usepackage{ifthen}
\usepackage{comment}
\newboolean{arxivversion}
\setboolean{arxivversion}{true}
\newcommand{\arxivexcl}[2]{\ifthenelse{\boolean{arxivversion}}{#1}{#2}}

\newboolean{confversion}
\setboolean{confversion}{false}
\ifthenelse{\boolean{confversion}}{\includecomment{confenv}}{\excludecomment{confenv}}
\newcommand{\confcmt}[1]{\ifthenelse{\boolean{confversion}}{#1}{}}
\newboolean{fullversion}
\setboolean{fullversion}{true}
\ifthenelse{\boolean{fullversion}}{\includecomment{fullenv}}{\excludecomment{fullenv}}
\newcommand{\fullcmt}[1]{\ifthenelse{\boolean{fullversion}}{#1}{}}
\newcommand{\fig}[6]{
  \begin{figure}[#1] % options, h, t, b, p
    \centering
    \begin{minipage}{#2\linewidth}
      \centering
      \arxivexcl{\includegraphics[#3]{#4}}
      {\includegraphics[#3]{figure/#4}}
      \caption{#6}
      \label{#5}
    \end{minipage}
  \end{figure}
}

\usepackage{pifont}         % Symbols in the todo's
\newcommand{\floor}[1]{\lfloor #1 \rfloor}

\renewcommand{\(}{\left (}
\renewcommand{\)}{\right )}
\let\mylceil=\lceil   \renewcommand{\lceil}{\left\mylceil}
\let\myrceil=\rceil   \renewcommand{\rceil}{\right\myrceil}
\let\mylfloor=\lfloor \renewcommand{\lfloor}{\left\mylfloor} 
\let\myrfloor=\rfloor \renewcommand{\rfloor}{\right\myrfloor}

\newcommand{\bigO}{\mathcal{O}}

\newcommand{\iref}[1]{I.\ref{#1}} % Invariant reference
\newcommand{\oref}[1]{O.\ref{#1}} % Observation reference
\newcommand{\op}[1]{{\rm\textsf{#1}}}
\newcommand{\slack}{\textsf{slack}}

\newcommand{\pred}{\textsf{p}}
\renewcommand{\succ}{\textsf{s}}
\newcommand{\GIL}{\textsf{GIL}}
\newcommand{\GIR}{\textsf{GIR}}
\newcommand{\FGL}{\textsf{FGL}}
\newcommand{\FGR}{\textsf{FGR}}
\newcommand{\ws}[1]{\ell_{#1}}

\newtheorem{thm}{Theorem}
\newtheorem{lem}{Lemma}
\begin{fullenv}
\title{Cache-Oblivious Implicit Predecessor Dictionaries with the Working-Set
Property}
\titlerunning{Cache-Oblivious Implicit Predecessor Dictionaries with the
Working-Set Property}
\end{fullenv}
\begin{confenv}
\title{Cache-Oblivious Implicit Predecessor Dictionaries with the Working-Set
Property\footnote{This is an extended abstract, the full paper is available at \url{http://arxiv.org/abs/1112.5472
}}}
\titlerunning{Cache-Oblivious Implicit Predecessor Dictionaries with the
Working-Set Property}
\end{confenv}

\author[1]{Gerth St{\o}lting Brodal}
\author[1]{Casper Kejlberg-Rasmussen}
\affil[1]{MADALGO\footnote{Center for Massive Data
Algorithmics, a Center of the Danish National Research Foundation}, Department of Computer
  Science, Aarhus University, Denmark} 
\authorrunning{G. S. Brodal and C. Kejlberg-Rasmussen}
\begin{confenv}
\Copyright[by-nd] %choose "by", "by-sa", "by-nd" or "by-nc-nd"
          {Gerth St{\o}lting Brodal and Casper Kejlberg-Rasmussen}
\end{confenv}
\begin{fullenv}
\Copyright[by] %choose "by", "by-sa", "by-nd" or "by-nc-nd"
          {Gerth St{\o}lting Brodal and Casper Kejlberg-Rasmussen}
\EventShortName{STACS 2012}
\end{fullenv}

\begin{confenv}
\serieslogo{}%please provide filename (without suffix)
\volumeinfo%(easychair interface)
  {Billy Editor, Bill Editors}% editors
  {2}% number of editors: 1, 2, ....
  {Conference title on which this volume is based on}% event
  {1}% volume
  {1}% issue
  {1}% starting page number
\EventShortName{}
\DOI{10.4230/LIPIcs.xxx.yyy.p}% to be completed by the volume editor
\end{confenv}

\confcmt{\bibliographystyle{abbrv}}
\fullcmt{\bibliographystyle{plain}}

\begin{document}
%%%%%%%%%%%%%%%%%%%%%%%%%%%%%%%%%%%%%%%%%%%%%%%%%%%%%%%%%%%%%%%%%%%%%%%%%%%%%%%%
\maketitle
%%%%%%%%%%%%%%%%%%%%%%%%%%%%%%%%%%%%%%%%%%%%%%%%%%%%%%%%%%%%%%%%%%%%%%%%%%%%%%%%

\begin{abstract}
In this paper we present an implicit dynamic dictionary with the working-set
property, supporting \op{insert}($e$) and \op{delete}($e$) in $\bigO(\log n)$
time, \op{predecessor}($e$) in $\bigO(\log \ws{\pred(e)})$ time,
\op{successor}($e$) in $\bigO(\log \ws{\succ(e)})$ time and \op{search}($e$) in
$\bigO(\log \min(\ws{\pred(e)},\ws{e}, \ws{\succ(e)}))$ time, where $n$ is the
number of elements stored in the dictionary, $\ws{e}$ is the number of distinct
elements searched for since element~$e$ was last searched for and $\pred(e)$
and $\succ(e)$ are the predecessor and successor of $e$, respectively. The
time-bounds are all worst-case. The dictionary stores the elements in an array
of size~$n$ using \emph{no} additional space. In the cache-oblivious model the
$\log$ is base $B$ and the cache-obliviousness is due to our black box use of
an existing cache-oblivious implicit dictionary. This is the first implicit
dictionary supporting predecessor and successor searches in the working-set
bound. Previous implicit structures required $\bigO(\log n)$ time.
\end{abstract}

%\begin{keywords}
%working-set property, dictionary, implicit, cache-oblivious, worst-case,
%external memory, I/O efficient
%\end{keywords}

%\thispagestyle{empty}
%\setcounter{page}{0}
%\clearpage

%%%%%%%%%%%%%%%%%%%%%%%%%%%%%%%%%%%%%%%%%%%%%%%%%%%%%%%%%%%%%%%%%%%%%%%%%%%%%%%%
\section{Introduction} \label{sec:introduction}
%%%%%%%%%%%%%%%%%%%%%%%%%%%%%%%%%%%%%%%%%%%%%%%%%%%%%%%%%%%%%%%%%%%%%%%%%%%%%%%%

In this paper we consider the problem of maintaining a cache-oblivious implicit
dictionary~\cite{MS80} with the working-set property over a dynamically
changing set $P$ of $|P|=n$ distinct and totally ordered elements. We define
the \emph{working-set number} of an element $e \in P$ to be $\ws{e} = |\{e' \in
P \mid$ we have searched for $e'$ after we last searched for $e\}|$.  An
implicit dictionary maintains $n$ distinct keys without using any other space
than that of the $n$ keys, i.e. the data structure is encoded by permuting the
$n$ elements. The fundamental trick in the implicit model, \cite{M86}, is to
encode a bit using two distinct elements $x$ and $y$: if $\min(x,y)$ is before
$\max(x,y)$ then $x$ and $y$ encode a 0 bit, else they encode a 1 bit. This can
then be used to encode $l$ bits using $2l$ elements. The implicit model is a
restricted version of the unit cost RAM model with a word size of $\bigO(\log
n)$. The restrictions are that between operations we are only allowed to use an
array of the $n$ input elements to store our data structures by permuting the
input elements, i.e., there can be used \emph{no} additional space between
operations. In operations we are allowed to use $\bigO(1)$ extra words.
Furthermore we assume that the number of elements $n$ in the dictionary is
externally maintained. Our structure will support the following operations:
\begin{itemize}
  \item \textsf{Search$(e)$} determines if $e$ is in the dictionary, if so
  its working-set number is set to $0$.
  
  \item \textsf{Predecessor$(e)$} will find $\max\{e' \in P \cup \{-\infty\}
  \mid e' < e\}$, without changing any working-set numbers.
  
  \item \textsf{Successor$(e)$} will find $\min\{e' \in P \cup \{\infty\} \mid
  e < e'\}$, without changing any working-set numbers.

  \item \textsf{Insert$(e)$} inserts $e$ into the dictionary with at working-set
  number of $0$, all other working-set numbers are increased by one.
  
  \item \textsf{Delete$(e)$} deletes $e$ from the dictionary, and does not
  change the working-set number of any element.
\end{itemize}
\fullcmt{There are numerous data structures and algorithms in the implicit
model which range from binary heaps \cite{W64} to in-place 3-D convex hull
algorithms \cite{CC09}. }There has been a continuous development of implicit
dictionaries, the first milestone was the implicit AVL-tree \cite{M86} having
bounds of $\bigO(\log^2 n)$. The second milestone was the implicit B-tree
\cite{FGMP02} having bounds of $\bigO(\log^2 n / \log \log n)$ the third was
the flat implicit tree \cite{FG06} obtaining $\bigO(\log n)$ worst-case time
for searching and amortized bounds for updates. The fourth milestone is the
optimal implicit dictionary \cite{FG03} obtaining worst-case $\bigO(\log n)$
for search, update, predecessor and successor.

Numerous non-implicit dictionaries attain the working-set property; splay trees
\cite{ST85}, skip list variants~\cite{BDL08}, the working-set structure in
\cite{I01}, and two structures presented in \cite{BHM09}. All achieve the
property in the amortized, expected or worst-case sense. The unified access
bound, which is achieved in \cite{BCDI07}, even combines the working-set
property with finger search. In finger search we have a finger located on an
element $f$ and the search cost of finding say element $e$ is a function of
$d(f,e)$ which is the rank distance between elements $f$ and $e$. The unified
bound combines these two to obtain a bound of $\bigO(\min_{e \in
P}\{\log(\ws{e} + d(e,f) + 2)\})$. Table \ref{tab:previousresults} gives an
overview of previous results, and our contribution.
\begin{table}[tb]
  \begin{center}
    \begin{tabular}{@{\extracolsep{0pt}}l@{\extracolsep{0pt}}|@{\extracolsep{0pt}}c@{\extracolsep{0pt}}|@{\extracolsep{0pt}}l@{\extracolsep{0pt}}|@{\extracolsep{0pt}}l@{\extracolsep{0pt}}|@{\extracolsep{0pt}}l@{\extracolsep{0pt}}|@{\extracolsep{0pt}}l@{\extracolsep{0pt}}}
        ~\textbf{Ref.} ~&~ \parbox{0.34in}{\textbf{WS prop.}} ~&~
        \parbox{0.8in}{\textbf{Insert/ Delete$(e)$}} ~&~ \textbf{Search$(e)$}
        ~&~ \parbox{0.8in}{\textbf{Pred$(e)$/ Succ$(e)$}} ~&~
        \parbox{0.75in}{\textbf{Additional words}} \\
      \hline
        ~\cite{M86} ~&~ -- ~&~ $\bigO(\log^2 n)$ ~&~ $\bigO(\log^2 n)$ ~&~ --
        ~&~ None \\

        ~\cite{FGMP02} ~&~ -- ~&~ $\bigO\(\frac{\log^2 n}{\log \log n}\)$ ~&~
        $\bigO\(\frac{\log^2 n}{\log \log n}\)$ ~&~ -- ~&~ None \\

        ~\cite{FG06} ~&~ -- ~&~ $\bigO(\log n)$ amor. ~&~ $\bigO(\log n)$ ~&~
        $\bigO(\log n)$ ~&~ None \\

        ~\cite{FG03} ~&~ -- ~&~ $\bigO(\log n)$ ~&~ $\bigO(\log n)$ ~&~
        $\bigO(\log n)$ ~&~ None \\

        ~\cite{I01} ~&~ + ~&~ $\bigO(\log n)$ ~&~ $\bigO(\log \ws{e})$ ~&~
        $\bigO(\log \ws{e^*})$ ~&~ $\bigO(n)$ \\

        ~\cite[Sec. 2]{BHM09} ~&~ + ~&~ $\bigO(\log n)$ ~&~ $\bigO(\log
        \ws{e})$ exp.  ~&~ $\bigO(\log n)$ ~&~ $\bigO(\log\log n)$ \\

        ~\cite[Sec. 3]{BHM09} ~&~ + ~&~ $\bigO(\log n)$ ~&~ $\bigO(\log
        \ws{e})$ exp.  ~&~ $\bigO(\log \ws{e^*})$ exp. ~&~ $\bigO(\sqrt{n})$ \\

        ~\cite{BKT10}~&~ + ~&~ $\bigO(\log n)$ ~&~ $\bigO(\log \ws{e})$ ~&~
        $\bigO(\log n)$ ~&~ None \\
      \hline
        ~This paper\hspace{0.15cm}~&~ + ~&~ $\bigO(\log n)$ ~&~ $\bigO(\log
        \min(\ws{\pred(e)},\ws{\succ(e)},\ws{e}))$ ~&~ $\bigO(\log \ws{e^*})$
        ~&~ None \\
    \end{tabular}
  \end{center}
  \caption{The operation time and space overhead of important structures for
  the dictionary problem. Here $e^*$ is the predecessor or successor in the
  given context. In a search for an element $e$ that is not present in the
  dictionary $\ws{e}$ is $n$.}
  \label{tab:previousresults}
\end{table}

The dictionary in \cite{FG03} is, in addition to being implicit, also designed
for the cache-oblivious model \cite{FLPR99}, where all the operations imply
$\bigO(\log_B n)$ cache-misses. Here $B$ is the cache-line length which is
unknown to the algorithm. The cache-oblivious property also carries over into
our dictionary. Our structure combines the two worlds of implicit dictionaries
and dictionaries with the working-set property to obtain the first implicit
dictionary with the working-set property supporting search, predecessor and
successor queries in the working-set bound. The result of this paper is
summarized in Theorem \ref{thm:results}.
\begin{thm} \label{thm:results}
  \sloppypar{There exists a cache-oblivious implicit dynamic dictionary with
  the working-set property that supports the operations insert and delete in
  time $\bigO(\log n)$ and $\bigO(\log_B n)$ cache-misses, search, predecessor
  and successor in time $\bigO(\log \min(\ws{\pred(e)},\ws{e},
  \ws{\succ(e)}))$, $\bigO(\log \ws{\pred(e)})$ and $\bigO(\log
  \ws{\succ(e)})$, and cache-misses $\bigO(\log_B \min(\ws{\pred(e)},\ws{e},
  \ws{\succ(e)}))$, $\bigO(\log_B \ws{\pred(e)})$ and $\bigO(\log_B
  \ws{\succ(e)})$, respectively, where $\pred(e)$ and $\succ(e)$ are the
  predecessor and successor of $e$, respectively.}
\end{thm}
Similarly to previous work \cite{BCDI07,BKT10} we partition the dictionary
elements into $\bigO(\log \log n)$ blocks $B_0,\ldots,B_m$, of double
exponential increasing sizes, where $B_0$ stores the most recently accessed
elements. The structure in \cite{BKT10} supports predecessors and successors
queries, but there is no way of knowing if an element is actually the
predecessor or successor, without querying all blocks, which results in
$\bigO(\log n)$ time bounds. We solve this problem by introducing the notion of
\emph{intervals} and particularly a dynamic implicit representation of these.
We represent the whole interval $[\min(P);\max(P)]$ by a set of disjoint
intervals spread across the different blocks. Any point that intersects an
interval in block $B_i$ will lie in block $B_i$ and have a working-set number
of at least $2^{2^i}$. This way when we search for the predecessor or successor
of an element and hit an interval, then no more points can be contained in the
interval in higher blocks, and we can avoid looking at these, which give
working-set bounds for the search, predecessor and successor queries.

%%%%%%%%%%%%%%%%%%%%%%%%%%%%%%%%%%%%%%%%%%%%%%%%%%%%%%%%%%%%%%%%%%%%%%%%%%%%%%%%
\section{Data structure}
%%%%%%%%%%%%%%%%%%%%%%%%%%%%%%%%%%%%%%%%%%%%%%%%%%%%%%%%%%%%%%%%%%%%%%%%%%%%%%%%

We now describe our data structure and its invariants. We will use the moveable
dictionary from \cite{BKT10} as a black box. The dictionary over a point set
$S$ is laid out in the memory addresses $[i;j]$. It supports the following
operations in $\bigO(\log n')$ time and $\bigO(\log_B n')$ cache-misses, where
$n'=j-i+1$:
\begin{itemize}
  \item \textsf{Insert-left$(e)$} inserts $e$ into $S$ which is now laid out in
  the addresses $[i-1;j]$.

  \item \textsf{Insert-right$(e)$} inserts $e$ into $S$ which is now laid out
  in the addresses $[i;j+1]$.

  \item \textsf{Delete-left$(e)$} deletes $e$ from $S$ which is now laid out in
  the addresses $[i+1;j]$.

  \item \textsf{Delete-right$(e)$} deletes $e$ from $S$ which is now laid out
  in the addresses $[i;j-1]$.

  \item \textsf{Search$(e)$} determines if $e \in S$, if so the address of
  element $e$ is returned.

  \item \textsf{Predecessor$(e)$} returns the address of the element $\max\{e'
  \in S \mid e' < e\}$ or that no such element exists.

  \item \textsf{Successor$(e)$} returns the address of the element $\min\{e'
  \in S \mid e < e'\}$ or that no such element exists.
\end{itemize}
From these operations we notice that we can move the moveable dictionary, say
left, by performing a delete-right operation for an arbitrary element and
re-inserting the element again by an insert-left operation. Similarly we can
also move the dictionary one position to the right.

Our structure consists of $m = \Theta(\log \log n)$ blocks $B_0, \ldots, B_m$,
each block $B_i$ is of size $\bigO(2^{2^{i+k}})$, where $k$ is a constant.
Elements in $B_i$ have a working-set number of at least $2^{2^{i+k-1}}$. The
block $B_i$ consists of an array~$D_i$ of $w_i = d\cdot 2^{i+k}$ elements,
where $d$ is a constant, and moveable dictionaries $A_i, R_i, W_i, H_i, C_i$
and $G_i$, for $i=0,\ldots,m-1$, see Figure~\ref{fig:memoryOverview}. For
block~$B_m$ we only have $D_m$ if $|B_m \backslash \{\min(P),\max(P)\}|
\leq w_m$, otherwise we have the same structures as for the other blocks. We use
the block $D_i$ to encode the sizes of the movable dictionaries $A_i, R_i, W_i,
H_i, C_i$ and $G_i$ so that we can locate them.  Discussion of further details
of the memory layout is postponed to Section \ref{sec:memoryManagement}.

\fig{tb}{}{}{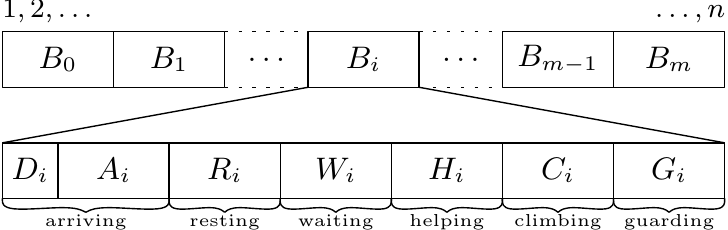}{fig:memoryOverview}{Overview of how the
working set dictionary is laid out in memory. The dictionary grows and shrinks
to the right when elements are inserted and deleted.}

We call elements in the structures $D_i$ and $A_i$ for \emph{arriving} points,
and when making a non-arriving point arriving, we will put it into $A_i$ unless
specified otherwise. We call elements in $R_i$ for \emph{resting} points,
elements in $W_i$ for \emph{waiting} points, elements in $H_i$ for
\emph{helping} points, elements in $C_i$ for \emph{climbing} points and
elements in $G_i$ for \emph{guarding} points.

Crucial to our data structure is the partitioning of $[\min(P);\max(P)]$ into
\emph{intervals}. Each interval is assigned to a \emph{level} and level $i$
corresponds to block $B_i$. Consider an interval lying at level $i$. The
endpoints $e_1$ and $e_2$ will be guarding points stored at level $0,\ldots,i$.
All points inside of this interval will lie in level $i$ and cannot be guarding
points, i.e. $]e_1;e_2[ \cap (\bigcup_{j \neq i}{B_j \cup G_i}) = \emptyset$.
We do not allow intervals defined by two consecutive guarding points to be
empty, they must contain at least one non-guarding point. We also require
$\min(P)$ and $\max(P)$ to be guarding points in $G_0$ at level $0$, but they
are special as they do not define intervals to their left and right,
respectively. A query considers $B_0,B_1,\ldots$ until $B_i$ where the query is
found to be in a level~$i$ interval where the answer is guaranteed to have been
found in blocks $B_0,\ldots,B_i$.

The basic idea of our construction is the following. When searching for an
element it is moved to level $0$. This can cause block overflows (see
invariants \iref{it:encode}--\iref{it:slack} in Section \ref{sec:invariants}),
which are handled as follows. The arriving points in level $i$ have just
entered from level $i-1$, and when there are $2^{2^{i+k}}$ of them in $A_i$
they become resting. The resting points need to charge up their working-set
number before they can begin their journey to level $i+1$. They are charged up
when there have come $2^{2^{i+k}}$ further arriving points to level~$i$, then
the resting points become waiting points. Waiting points have a high enough
working-set number to begin the journey to level $i+1$, but they need to wait
for enough points to group up so that they can start the journey. When a
waiting point is picked to start its journey to level $i+1$ it becomes a
helping or climbing point, and every time enough helping points have grouped
up, i.e.\ there is at least $c=5$ consecutive of them, then they become
climbing points and are ready to go to level $i+1$. The climbing points will
then incrementally be going to level $i+1$.\fullcmt{ See Figure
\ref{fig:overview} for an example of the structure of the intervals.}

\fullcmt{
\fig{tb}{}{}{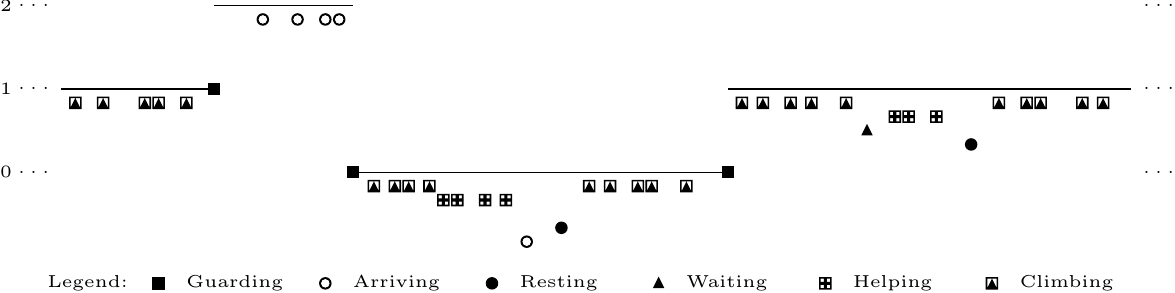}{fig:overview}{The structure of the
levels for a dictionary. The levels are indicated to the left.}}

\subsection{Notation}

Before we introduce the invariants we need to define some notation. For a subset
$S \subseteq P$, we define $\pred_S(e) = \max\{s \in S \cup \{-\infty\} \mid s <
e\}$ and $\succ_S(e) = \min\{s \in S \cup \{\infty\} \mid e < s\}$. When we
write $S_{\leq i}$ we mean $\bigcup_{j=0}^i{S_j}$ where $S_j \subseteq P$ for
$j=0,\ldots,i$.

For $S \subseteq P$, we define $\GIL_S(e) = S \cap ]\pred_{P\backslash S}(e);
e[$ to be the Group of Immediate Left points of $e$ in $S$ which does not have
any other point of $P\backslash S$ in between them\fullcmt{, see Figure
\ref{fig:notation}}. Similarly we define $\GIR_S(e) = S \cap
]e;\succ_{P\backslash S}(e)[$ to the right of~$e$. We will notice that we will
never find all points of $\GIL_S(e)$ unless $|\GIL_S(e)| < c$, the same applies
for $\GIR_S(e)$. For $S \subseteq P$, we define $\FGL_S(e) = S \cap
]\pred_{P\backslash S}(\pred_S(e)) ; \pred_S(e)]$ to be the First Group of
points from $S$ Left of $e$, i.e.\ the group does not have any points of
$P\backslash S$ in between its points\fullcmt{, see Figure \ref{fig:notation}}.
Similarly we define $\FGR_S(e) = S \cap [\succ_S(e) ; \succ_{P\backslash
S}(\succ_S(e))[$. We will notice that we will never find all points of
$\FGL_S(e)$ unless $|\FGL_S(e)| < c$, the same applies for $\FGR_S(e)$.

\fullcmt{
\fig{b}{0.99}{}{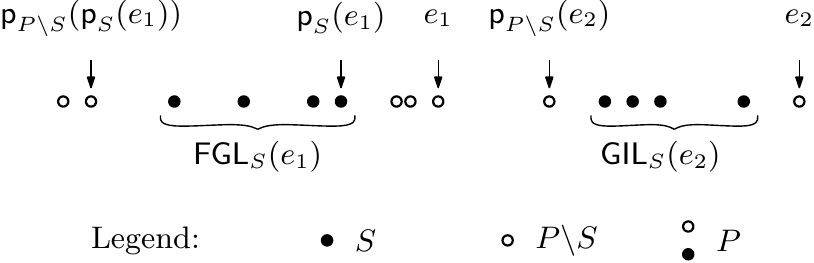}{fig:notation}{Here is a illustration of $\FGL$ and
$\GIL$. Notice that $\GIL_S(e_1) = \emptyset$ whereas $\FGL_S(e_1) \neq
\emptyset$.}}

We will sometimes use the phrasings \emph{a group of points} or \emph{$e$'s
group of points}. This refers to a group of points of the same type, i.e.
arriving, resting, etc., and with no other types of points in between them.
Later we will need to move elements around between the structures $D_i$, $A_i$,
$R_i$, $W_i$, $H_i$, $C_i$ and $G_i$. For this we have the notation $X
\stackrel{h}{\rightarrow} Y$, meaning that we move $h$ arbitrary points from
$X$ into $Y$, where $X$ and $Y$ can be one of $D_i$, $A_i$, $R_i$, $W_i$,
$H_i$, $C_i$ and $G_i$ for any $i$.

When we describe the intervals we let $]a;b]$ be an interval from $a$ to $b$
that is open at $a$ and closed at~$b$. We let $(a;b)$ be an interval from $a$
to $b$ that can be open or closed at $a$ and $b$. We use this notation when we
do not care if $a$ and $b$ are open or closed. In the methods updating the
intervals we will sometimes branch depending on which type an interval is. For
clarity we will explain how to determine this given the level $i$ of the
interval and its two endpoints $e_1$ and $e_2$. The interval $(e_1;e_2)$ is of
type $[e_1;e_2)$ if $e_1 \in G_i$, else $e_1 \in G_{\leq i-1}$ and the interval
is of type $]e_1;e_2)$. This is symmetric for the other endpoint~$e_2$.

%%%%%%%%%%%%%%%%%%%%%%%%%%%%%%%%%%%%%%%%%%%%%%%%%%%%%%%%%%%%%%%%%%%%%%%%%%%%%%%%
\subsection{Invariants} \label{sec:invariants}
%%%%%%%%%%%%%%%%%%%%%%%%%%%%%%%%%%%%%%%%%%%%%%%%%%%%%%%%%%%%%%%%%%%%%%%%%%%%%%%%

We will now define the invariants which will help us define and prove
correctness of our interface operations: \op{insert}$(e)$, \op{delete}$(e)$,
\op{search}$(e)$, \op{predecessor}$(e)$ and \op{successor}$(e)$. We maintain
the following invariants which uniquely determine the intervals\footnote{We
assume that $|P|=n \geq 2$ at all times if this is not the case we only store
$G_0$ which contains a single element and we ignore all invariants.}:
\begin{enumerate}[{I.}1]
  \item \label{it:intervals} A guarding point is part of the definition of at
  most two intervals\footnote{Only the smallest and largest guarding points
  will not participate in the definition of two intervals, all other guarding
  points will.}, one to the left at level $i$ and/or one to the right at level
  $j$, where $i\neq j$. The guarding point $e$ lies at level $\min(i,j)$. The
  interval at level $\min(i,j)$ is closed at $e$, and the interval at level
  $\max(i,j)$ is open at $e$. We also require that $\min(P)$ and $\max(P)$ are
  guarding points stored in $G_0$, but they do not define an interval to their
  left and right, respectively, and the intervals they help define are open in
  the end they define. A non-guarding point intersecting an interval at level
  $i$, lies in level $i$. Each interval contains at least one non-guarding
  point. The union of all intervals give $]\min(P);\max(P)[$.

  \item \label{it:cgroups} Any climbing point, which lies in an interval with
  other non-climbing points, is part of a group of at least $c$ points. In
  intervals of type $[e_1;e_2]$ which only contain climbing points, we allow
  there to be less than $c$ of them.

  \item \label{it:hneighbors} Any helping point is part of a group of size at
  most $c-1$. A helping point cannot have a climbing point as a predecessor or
  successor. An interval of type $[e_1;e_2]$ cannot contain only helping points.
\end{enumerate}

\noindent We maintain the following invariants for the working-set numbers:
\begin{enumerate}[{I.}1]
  \setcounter{enumi}{3}
  \item \label{it:wsvalue} Each arriving point in $D_i$ and $A_i$ has a working
  set value of at least $2^{2^{i-1+k}}$, arriving points in $D_0$ and $A_0$ have
  a working-set value of at least $0$. Each resting point in $R_i$ will have a
  working-set value of at least $2^{2^{i-1+k}}+|A_i|$, resting points in $R_0$
  have a working-set value of at least $|A_0|$. Each waiting, helping or
  climbing point in $W_i,H_i$ and $C_i$, respectively, will have a working-set
  value of at least $2^{2^{i+k}}$. Each guarding point in $G_i$, who's left
  interval lies at level $i$ and right interval lies at level $j$, has a working
  set value of at least $2^{2^{\max(i,j)-1+k}}$.
\end{enumerate}

\noindent We maintain the following invariants for the size of each block and
their components:
\begin{enumerate}[{I.}1]
  \setcounter{enumi}{4}
  \item \label{it:encode} $|D_0| = \min(|B_0|-2,w_0)$ and $|D_i| =
  \min(|B_i|,w_i)$ for $i=1,\ldots,m$.

  \item \label{it:resting} $|R_i| \leq 2^{2^{i+k}}$ and $|W_i|+|H_i|+|C_i| \neq
  0 \Rightarrow |R_i| = 2^{2^{i+k}}$ for $i = 0,\ldots,m$.

  \item \label{it:arrivingwaiting} $|A_i|+|W_i| = 2^{2^{i+k}}$ for
  $i=0,\ldots,m-1$, and $|A_m|+|W_m| \leq 2^{2^{m+k}}$.

  \item \label{it:arriving} $|A_i| < 2^{2^{i+k}}$ for $i=0,\ldots,m$.

  \item \label{it:slack} $|H_i| + |C_i| = 4c 2^{2^{i+k}} + c_i$, where $c_i \in
  [-c;c]$, for $i=0,\ldots,m-1$.
\end{enumerate}

\noindent From the above invariants we have the following observation:
\begin{enumerate}[{O.}1]
  \item \label{it:limitedguarding} From \iref{it:intervals} all points in $G_i$
  are endpoints of intervals in level $i$, and each interval has at most two
  endpoints. Hence for $i=0,\ldots,m$ we have that
  \[
    |G_i| \leq 2(|D_i| + |A_i| + |R_i| + |W_i| +
  |H_i| + |C_i|) \stackrel{(*)}{\leq} (4+2d+8c) \cdot 2^{2^{i+k}} + 2c\;,
  \]
  where we in $(*)$ we have used \iref{it:encode}, \iref{it:resting},
  \iref{it:arrivingwaiting} and \iref{it:slack}.
\end{enumerate}

\noindent From \iref{it:intervals} we have the following lemma.
\begin{lem} \label{lem:intervals}
  Let $e$ be an element, $e_1 = \pred_{G_{\leq i}}(e)$, $e_2 = \succ_{G_{\leq
  i}}(e)$ and $i$ be the smallest integer for which $I(e_1,e_2,i)=]e_1;e_2[
  \cap \bigcup_{j=0}^i{B_j} \neq \emptyset$. Then 1) $(e_1;e_2)$ is an interval
  at level $i$ if $e$ is non-guarding and 2) $(e_1;e)$ or $(e;e_2)$ is an
  interval at level $i$ if $e$ is guarding.
\end{lem}
\fullcmt{
\begin{proof}
  Assume that $i$ is the minimal $i$ that fulfills $I(e_1,e_2,i) \neq
  \emptyset$, where $e_1 = \pred_{G_{\leq i}}(e)$ and $e_2 = \succ_{G_{\leq
  i}}(e)$. We will have two cases depending on if $e$ is guarding or not.
  
  Lets first handle case 2) where $e$ is guarding and hence in the dictionary:
  Since $e$ is in the dictionary and $e_1 < e < e_2$ we have from the
  minimality of $i$ that $e$ lies in level $i$, and from \iref{it:intervals}
  $e$ is then part of an interval lying in level $i$ either to the left or to
  the right. Say $e$ is part of an interval to the left i.e.\ the interval
  $(e'_1;e)$. If $e_1 < e'_1$ then this would contradict that $e_1 =
  \pred_{G_{\leq i}}(e)$ hence $e'_1 \leq e_1$, but since $e'_1$ is the
  predecessor of $e$ we have that $e'_1=e_1$. So we know that $(e_1;e)$ defines
  an interval at level $i$. The argument for $(e;e_2)$ is symmetric.

  In the case 1) $e$ is non-guarding and $e$ may lie in the dictionary or not:
  Since $e_1 < e < e_2$ we have from the minimality of $i$ that $e$ lies in
  level $i$, hence from \iref{it:intervals} we have that the interval
  $(e_1;e_2)$ lies at level $i$.
\end{proof}}

%%%%%%%%%%%%%%%%%%%%%%%%%%%%%%%%%%%%%%%%%%%%%%%%%%%%%%%%%%%%%%%%%%%%%%%%%%%%%%%%
\subsection{Operations} \label{sec:operations}
%%%%%%%%%%%%%%%%%%%%%%%%%%%%%%%%%%%%%%%%%%%%%%%%%%%%%%%%%%%%%%%%%%%%%%%%%%%%%%%%

We will briefly give an overview of the helper operations and state their
requirements~\textit{(R)} and guarantees~\textit{(G)}, then we will describe
the helper and interface operations in details. Search$(e)$ uses the helper
operations as follows: when a search for element~$e$ is performed then the
level~$i$ where~$e$ lies is found using find, then~$e$ and~$\bigO(1)$ of its
surrounding elements are moved into level~$0$ by use of move-down while
maintaining \iref{it:intervals}--\iref{it:wsvalue}. Calls to fix for the levels
we have altered will ensure that \iref{it:encode}--\iref{it:arriving} will be
maintained, finally a call to rebalance-below$(i-1)$ will ensure that
\iref{it:slack} is maintained by use of shift-up$(j)$ which will take climbing
points from level~$j$ and make them arriving in level~$j+1$ for
$j=0,\ldots,i-1$. Insert$(e)$ uses find to find the level where $e$ intersects,
then it uses fix to ensure the size constraints and finally $e$ is moved to
level $0$ by use of search.

\begin{itemize}
  \item \textsf{Find$(e)$} - returns the level $i$ of the interval that $e$
  intersects along with $e$'s type and whatever $e$ is in the dictionary or
  not. \textit{[R\&G: \iref{it:intervals}--\iref{it:slack}]}

%  \item \textsf{Predecessor$(e)$ and successor$(e)$} - \textit{R\&G
%  \iref{it:intervals}--\iref{it:slack}:} returns the predecessor $\pred(e)$
%  (successor $\succ(e)$) of the element $e$ in the dictionary.

%  \item \textsf{Insert$(e)$} - \textit{R\&G
%  \iref{it:intervals}--\iref{it:slack}:} Inserts the element $e$ into the
%  dictionary with a working set value of $\ws{e} = 0$.

  \item \textsf{Fix$(i)$} - moves points around inside of $B_i$ to ensure the
  size invariants for each type of point. Fix$(i)$ might violate
  \iref{it:slack} for level $i$. \textit{[R:
  \iref{it:intervals}--\iref{it:wsvalue} and that there exist $\tilde{c}_1,
  \ldots, \tilde{c}_6$ such that $|D_i|+\tilde{c}_1, |A_i|+\tilde{c}_2,
  |R_i|+\tilde{c}_3, |W_i|+\tilde{c}_4, |H_i|+\tilde{c}_5, |C_i|+\tilde{c}_6$
  fulfill \iref{it:encode}--\iref{it:arriving}, where $|\tilde{c}_i| =
  \bigO(1)$ for $i=1,\ldots,6$. G: \iref{it:intervals}--\iref{it:arriving}].}

  \item \textsf{Shift-down$(i)$} - will move at least $1$ and at most $c$
  points from level $i$ into level $i-1$. \textit{[R:
  \iref{it:intervals}--\iref{it:arriving} and $|H_i| + |C_i| = 4c2^{2^{i+k}} +
  c'_i$, where $0 \leq c'_i = \bigO(1)$. G:
  \iref{it:intervals}--\iref{it:arriving}].}

  \item \textsf{Shift-up$(i)$} - will move at least $1$ and at most $c$ points
  from level $i$ into level $i+1$. \textit{[R:
  \iref{it:intervals}--\iref{it:arriving} and $|H_i|+|C_i| =
  4c2^{2^{i+k}}+c'_i$, where $c \leq c'_i = \bigO(1)$.  G:
  \iref{it:intervals}--\iref{it:arriving}].}

%  \item \textsf{Search$(e)$} \textit{R\&G
%  \iref{it:intervals}--\iref{it:slack}:} If $e$ is in the dictionary it gets a
%  working set number of $\ws{e} = 0$.

  \item \textsf{Move-down$(e,i,j,t_{\text{before}},t_{\text{after}})$} - If $e$
  is in the dictionary at level $i$ it is moved from level $i$ to level $j$,
  where $i \geq j$. The type $t_\text{before}$ is the type of $e$ before the
  move and $t_\text{after}$ is the type that $e$ should have after the move,
  unless $i=j$ in which case $e$ will be made arriving in level $j$.
  \textit{[R\&G: \iref{it:intervals}--\iref{it:arriving}].}

%  \item \textsf{Delete$(e)$}\textit{R\&G \iref{it:intervals}--\iref{it:slack}:}
%  will delete element $e$ from the dictionary without changing any other
%  elements working set number.

  \item \textsf{Rebalance-below$(i)$} - If any $c < c_l$ for $l=0,\ldots,i$
  rebalance-below$(i)$ will correct it so \iref{it:slack} will be fulfilled
  again for $l=0,\ldots,i$. \textit{[R: \iref{it:intervals}--\iref{it:arriving}
  and $\sum_{l=0}^{i}{\slack(c_l)} = \bigO(1)$, where
  \[
    \slack(c_l) = \left\{
    \begin{array}{cl}
      0 & \textit{if} \quad c_l \in [-c;c]\;, \\
      |c_l|-c & \textit{otherwise}\;.
    \end{array} \right.
  \]
  G: \iref{it:intervals}--\iref{it:slack}].}

  \item \textsf{Rebalance-above$(i)$} - If any $c_l < -c$ for $l=i,\ldots,m-1$
  rebalance-above$(i)$ will correct it so \iref{it:slack} will be fulfilled
  again for $l=i,\ldots,m-1$. \textit{[R:
  \iref{it:intervals}--\iref{it:arriving} and $\sum_{l=i}^{m-1}{\slack(c_l)} =
  \bigO(1)$. G: \iref{it:intervals}--\iref{it:slack}].}
\end{itemize}

\subparagraph*{Find$(e)$} We start at level $i=0$. If $e < \min(P)$ or $\max(P)
< e$ we return false and $0$. For each level we let $e_1 = \pred_{G_{\leq
i}}(e)$, $e_2 = \succ_{G_{\leq i}}(e)$, $p = \pred_{B_i \backslash G_i}(e)$ and
$s = \succ_{B_i \backslash G_i}(e)$. We find $p$ and $s$ by querying each of
the structures $D_i, A_i, R_i, W_i, H_i$ and $C_i$, we find $e_1$ and $e_2$ by
querying $G_i$ and comparing with the values of $e_1$ and $e_2$ from level
$i-1$. While $p < e_1$ and $e_2 < s$ we continue to the next level, that is we
increment $i$. Now outside the loop, if $e \in B_i$ we return~$i$, the type of
$e$ and the boolean true as we found $e$, else we return $i$ and false as we
did not find $e$.\fullcmt{ See Figure \ref{fig:find-pred-succ} for an example
of the execution.}

\subparagraph*{Predecessor$(e)$ (successor$(e)$)} We start at level $i=0$. If
$e < \min(P)$ then return $-\infty$ ($\min(P)$). If $\max(P) < e$ then return
$\max(P)$ ($\infty$). For each level we let $e_1 = \pred_{G_{\leq i}}(e)$, $p =
\pred_{B_i}(e)$, $e_2 = \succ_{G_{\leq i}}(e)$ and $s = \succ_{B_i}(e)$. While
$p < e_1$ and $e_2 < s$ we continue to the next level, that is we increment
$i$. When the loop breaks we return $\max(e_1,p)$ ($\min(s,e_2)$).\fullcmt{ See
Figure \ref{fig:find-pred-succ} for an example of the execution.}

\fullcmt{
\fig{tb}{0.99}{}{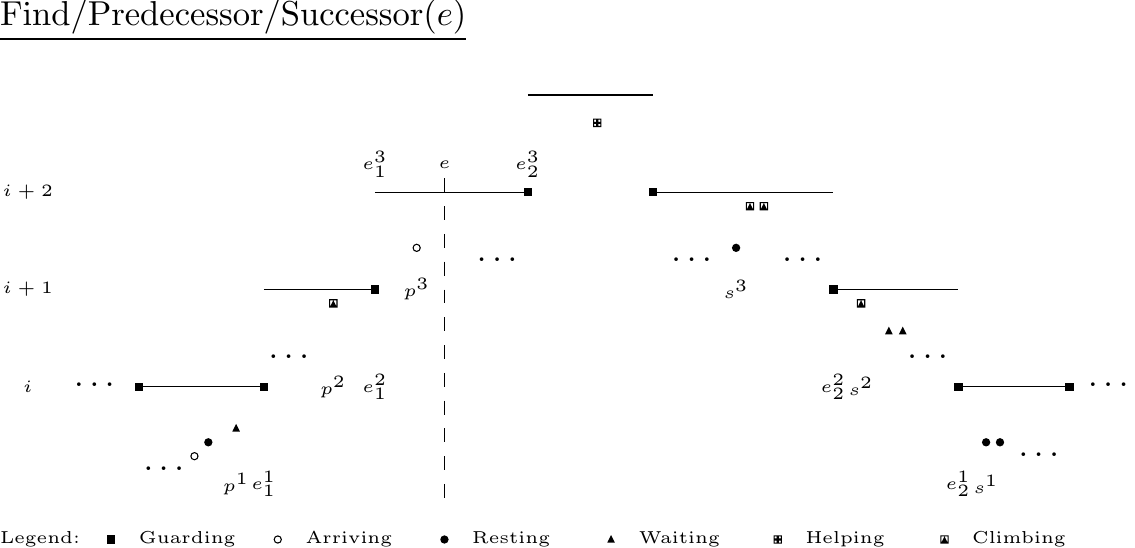}{fig:find-pred-succ}{The last three
iterations of the while-loop of find$(e)$, predecessor$(e)$ and
successor$(e)$.}}

\subparagraph*{Insert$(e)$} If $e < \min(P)$ we swap $e$ and $\min(P)$, call
fix$(0)$, rebalance-below$(m)$ and return. If $\max(P) < e$ we swap $e$ and
$\max(P)$, call fix$(0)$, rebalance-below$(m)$ and return.

Let $c_l = \GIL_{C_i}(e)$, $c_r = \GIR_{C_i}(e)$, $h_l = \GIL_{H_i}(e)$ and
$h_r = \GIR_{H_i}(e)$. We find the level $i$ of the interval $(e_1;e_2)$ which
$e$ intersects using find$(e)$.

If $e$ is already in the dictionary we give an error. If $|c_l| > 0$ or $|c_r|
> 0$ or $(e_1;e_2)$ is of type $[e_1;e_2]$ and does not contain non-climbing
points then insert $e$ as climbing at level $i$. Else if $|h_l| + 1 + |h_r|
\geq c$ then insert $e$ as climbing at level $i$ and make the points in $h_l$
and $h_r$ climbing at level $i$. Else insert $e$ as helping at level $i$.
Finally we call rebalance-below$(m)$ and then search$(e)$ to move $e$ from the
current level $i$ down to level $0$.

\subparagraph*{Search$(e)$} We first find $e$'s current level $i$ and its type
$t$, by a call to find$(e)$. If $e$ is in the dictionary then we call
move-down$(e,i,0,t, \text{arriving})$ which will move $e$ from level $i$ down
to level $0$ and make it arriving, while maintaining
\iref{it:intervals}--\iref{it:arriving}, but \iref{it:slack} might be broken so
we finally call rebalance-below$(i-1)$ to fix this.

\subparagraph*{Fix$(i)$} In the following we will be moving elements around
between $D_i$, $A_i$, $R_i$, $W_i$, $H_i$ and $C_i$. The moves $A_i \rightarrow
R_i$ and $R_i \rightarrow W_i$, i.e.\ between structures which are next to each
other in the memory layout, are simply performed by deleting an element from the
left structure and inserting it into the right structure. The moves $W_i
\rightarrow H_i \cup C_i$ and the other way around $H_i \cup C_i \rightarrow
W_i$ will be explained below.

If $|D_i| > w_i$ then perform $D_i \stackrel{h}{\rightarrow} A_i$ where $h =
|D_i| - w_i$. If $|D_i| < w_i$ and $|B_i \backslash \{\min(P),$ $\max(P)\}| >
|D_i|$ then perform $H_i \cup C_i \stackrel{h_1}{\rightarrow} W_i$, $W_i
\stackrel{h_2}{\rightarrow} R_i$, $R_i \stackrel{h_3}{\rightarrow} A_i$ and
$A_i \stackrel{h_4}{\rightarrow} D_i$ where $h_1 =
\min(w_i-|D_i|,|H_i|+|C_i|)$, $h_2 = \min(w_i-|D_i|, |W_i|+h_1)$, $h_3 =
\min(w_i-|D_i|, |R_i|+h_2)$ and $h_4 = \min(w_i-|D_i|, |A_i|+h_3)$.

If $|W_i|+|H_i|+|C_i| \neq 0$ and $|R_i| < 2^{2^{i+k}}$ then perform $H_i \cup
C_i \stackrel{h_1}{\rightarrow} W_i$ and $W_i \stackrel{h_2}{\rightarrow} R_i$
where $h_1 = \min(2^{2^{i+k}} - |R_i|, |H_i| + |C_i|)$ and $h_2 =
\min(2^{2^{i+k}} - |R_i|, |W_i|+h_1)$. If $|R_i| > 2^{2^{i+k}}$ then perform
$R_i \stackrel{h_1}{\rightarrow} A_i$ where $h_1=|R_i| - 2^{2^{i+k}}$.

If $i<m$ and $|A_i|+|W_i| < 2^{2^{i+k}}$ then perform $H_i \cup C_i
\stackrel{h_1}{\rightarrow} W_i$, where $h_1 = \min(2^{2^{i+k}} -
(|A_i|+|W_i|), |H_i| + |C_i|)$. If $|A_i|+|W_i| > 2^{2^{i+k}}$ then perform
$W_i \stackrel{h_1}{\rightarrow} H_i \cup C_i$ where $h_1 = \min(|A_i|+|W_i| -
2^{2^{i+k}}, |W_i|)$.

If $|A_i| \geq 2^{2^{i+k}}$ then let $h_1 = |A_i| - 2^{2^{i+k}}$, delete $W_i$
as it is empty and rename $R_i$ to $W_i$. Now move $h_1$ elements from $A_i$
into a new moveable dictionary $X$, rename $A_i$ to $R_i$, rename $X$ to $A_i$
and perform $W_i \stackrel{h_1}{\rightarrow} H_i \cup C_i$.  \\

\noindent \textsl{\textsf{Performing $W_i \rightarrow H_i \cup C_i$:}} Let $w =
\succ_{W_i}(-\infty)$, $c_l = \GIL_{C_i}(w)$, $c_r = \GIR_{C_i}(w)$, $h_l =
\GIL_{H_i}(w)$ and $h_r = \GIR_{H_i}(w)$. If $|c_l| > 0$ or $|c_r| > 0$ or
$(e_1;e_2)$ is of type $[e_1;e_2]$ and only contains climbing points then make
$w$ climbing at level $i$. Else if $|h_l| + 1 + |h_r| \geq c$ then make $h_l$,
$w$ and $h_r$ climbing at level $i$. Else make $w$ helping at level $i$. \\

\noindent \textsl{\textsf{Performing $H_i \cup C_i \rightarrow W_i$:}} Let $w$
be the minimum element of $\succ_{H_i}(-\infty)$ and $\succ_{C_i}(-\infty)$,
and let $c_r = \GIR_{C_i}(w)$. Make $w$ waiting at level $i$. If $w$ was
climbing and $|c_r| < c$ then make $c_r$ helping at level $i$.

% This will move 3-? elements down to level i-1 from level i
\subparagraph{Shift-down$(i)$} We move at least one element from level $i$ into
level $i-1$\fullcmt{, see Figure \ref{fig:find-pred-succ}}. If $|D_i| < w_i$
then we let $a$ be some element in $D_i$. If $|D_i| < |B_i|$ then: if $|A_i| =
0$ we perform\footnote{The move $H_i \cup C_i \stackrel{l}{\rightarrow} W_i$
will be performed the same way as we did it in fix.} $H_i \cup C_i
\stackrel{h_1}{\rightarrow} W_i$, $W_i \stackrel{h_2}{\rightarrow} R_i$ and
$R_i \rightarrow A_i$, where $h_1=\min(1,|H_i|+|C_i|)$ and $h_2 =
\min(1,|W_i|+h_1)$, now we know that $|A_i| > 0$ so let
$a=\succ_{A_i}(-\infty)$, i.e., $a$ is the leftmost arriving point in $A_i$ at
level~$i$. We call move-down$(a,i,i-1, \text{arriving}, \text{climbing})$.

% This will move 1-5? elements up to level i+1 from level i.
\subparagraph*{Shift-up$(i)$} Assume we are at level $i$, we want to move at
least one and at most $c$ arbitrary points from $B_i$ into $B_{i+1}$.
Let\fullcmt{\footnote{See the analysis in Section \ref{sec:analysis} for a
proof that $|C_i|>0$.}} $s_1=\succ_{C_i}(-\infty)$, $e_1=\pred_{G_{\leq
i}}(s_1)$ and $e_2=\succ_{G_{\leq i}}(s_1)$, and let $s_2=\succ_{C_i \cap
[e_1;e_2]}(s_1)$, $s_3=\succ_{C_i \cap [e_1;e_2]}(s_2)$, $s_4=\succ_{C_i \cap
[e_1;e_2]}(s_3)$ and $s_5=\succ_{C_i \cap [e_1;e_2]}(s_4)$, if they exist, also
let $c_r = \GIR_{C_i}(s_4)$ be the group of climbing elements to the immediate
right of $s_4$, if they exist\fullcmt{, see Figure \ref{fig:Intervals}}. We
will now move one or more climbing points from $B_i$ into $B_{i+1}$ where they
become arriving points. If $i=m-1$ or $i=m$ then we put arriving points into
$D_{i+1}$, which we might have to create, instead of $A_{i+1}$.

We now deal with the case where $(e_1;e_2)$ is of type $[e_1;e_2]$ and only
contains climbing points. Let $l$ be the level of $e_1$'s left interval, and
$r$ the level of $e_2$'s right interval, also let $c_I$ be the number of
climbing points in the interval. If $l=i+1$ we make $e_1$ arriving, else we
make it guarding, at level $i+1$. Make the points of $s_1,s_2,s_3$ and $s_4$
that exist arriving at level $i+1$. If $c_I \leq c$ then make $s_5$ arriving at
level $i+1$ if it exists, also if $r=i+1$ we make $e_2$ arriving, else we make
it guarding, at level $i+1$. Else make $s_5$ guarding at level~$i$.

We now deal with the cases where $(e_1;e_2)$ might contain non-climbing points.
If $\pred(s_1) = e_1$ we make $s_1$ and $s_2$ waiting and guarding at level
$i$, respectively, else we make $s_1$ guarding at level $i$ and $s_2$ arriving
at level $i+1$. Now in both cases we make $s_3$ arriving at level $i+1$ and
$s_4$ guarding at level $i$. If $\langle (s_4;e_2)$ is not of type $[s_4;e_2]$
or contains non-climbing points$\rangle$ and $|c_r| < c$, i.e. there are less
than $c$ consecutive climbing points to the right of $s_4$, then we make the
points $c_r$ helping at level $i$.

We have moved climbing points from $B_i$ into $B_{i+1}$, and made them
arriving. Finally we call fix$(i+1)$.

\fullcmt{
\fig{tb}{0.99}{}{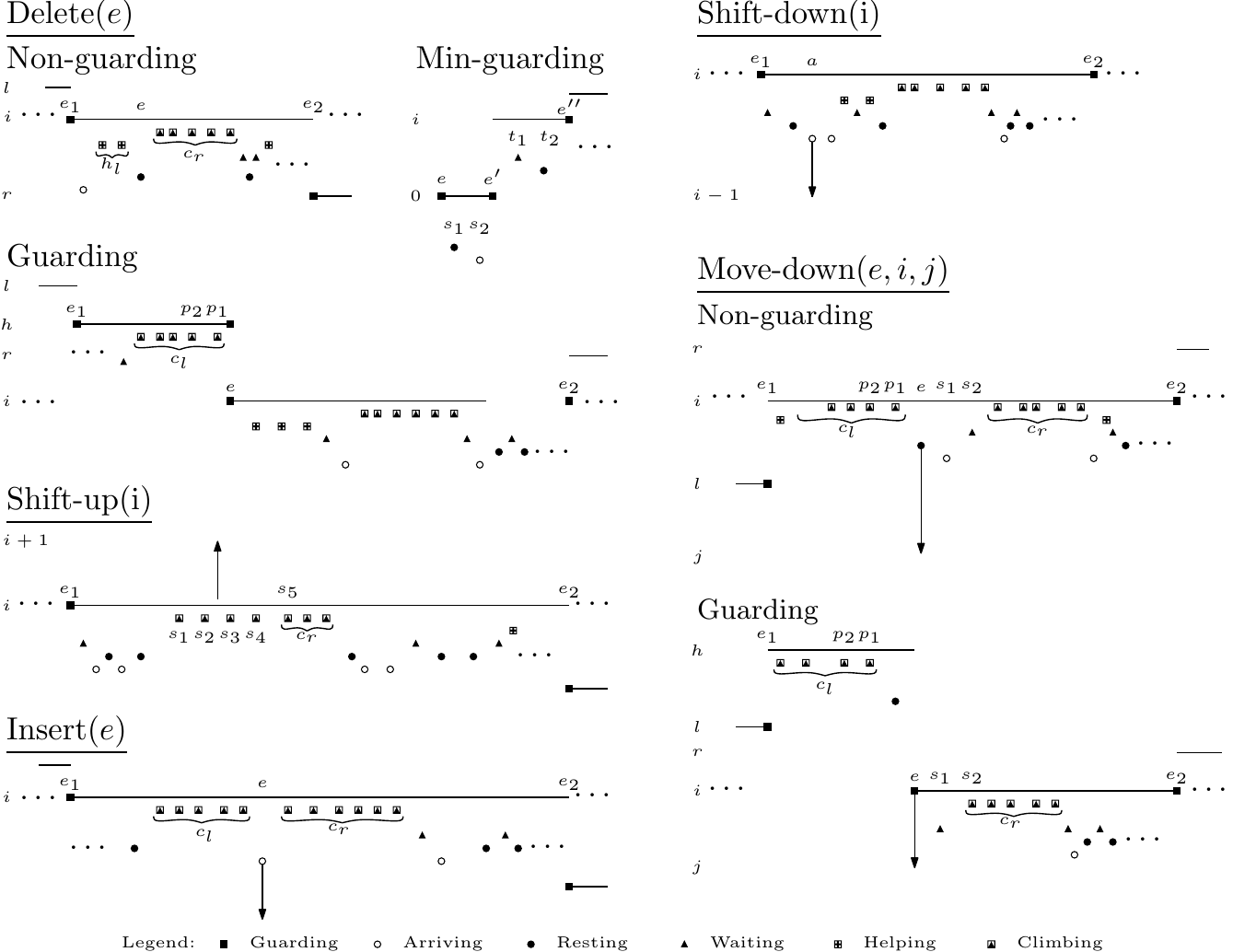}{fig:Intervals}{Here we see illustrations
of how we maintain the intervals when updating the intervals. These only show
single cases of each of the update methods many cases.}}

\subparagraph*{Move-down$(e,i,j,t_{\text{before}},t_{\text{after}})$}Depending
on the type $t_{\text{before}}$ of point $e$ we have different cases\fullcmt{,
see Figure \ref{fig:Intervals}}. \\

\noindent \textsl{\textsf{Non-guarding}} Let $e_1 = \pred_{G_{\leq i}}(e)$,
$e_2 = \succ_{G_{\leq i}}(e)$ and let $l$ be the level of the left interval of
$e_1$ and $r$ the level of the right interval of $e_2$. Also let $p_2 =
\pred_{B_i \backslash G_i \cap [e_1;e_2]}(p_1)$, $p_1 = \pred_{B_i \backslash
G_i \cap [e_1;e_2]}(e)$, $s_1 = \succ_{B_i \backslash G_i \cap [e_1;e_2]}(e)$
and $s_2 = \succ_{B_i \backslash G_i \cap [e_1;e_2]}(s_1)$, also let $c_l =
\FGL_{C_i \cap [e_1;e_2]}(e)$ be the elements in the first climbing group left
of $e$, likewise let $c_r = \FGR_{C_i \cap [e_1;e_2]}(e)$ be the elements in
the first climbing group right of $e$.

Case $i=j$: make $e$ arriving in level $j$, if $|c_l| < c$ then make the points
in $c_l$ helping at level $j$, if $|c_r| < c$ then make the points in $c_r$
helping at level $j$. Finally call fix$(j)$.

Case $i>j$: If both $p_2$ and $p_1$ exists we make $p_1$ guarding in level $j$
and let $e_1'$ denote~$p_1$, else if only $p_1$ exists we make $e_1$ guarding
at level $\min(l,j)$ and $p_1$ of type $t_{\text{after}}$ at level~$j$ and let
$e_1'$ denote $e_1$, else we make $e_1$ guarding in level $\min(l,j)$, and let
$e_1'$ denote $e_1$. If both $s_1$ and $s_2$ exists we make $s_1$ guarding at
level $j$, and let $e_2'$ denote $s_1$, else if only $s_1$ exists we make $s_1$
of type $t_{\text{after}}$ at level $j$ and make $e_2$ guarding at level
$\min(j,r)$ and let $e_2'$ denote $e_2$, else we make $e_2$ guarding at level
$\min(j,r)$ and let $e_2'$ denote $e_2$. Lastly we make $e$ of type
$t_{\text{after}}$ in level $j$. Now let $c_l'$ denote the elements of $c_l$
which we have not moved in the previous steps, likewise let $c_r'$ denote the
elements of $c_r$ which we have not moved. If $\langle (e_1;e_1']$ is not of
type $[e_1;e_1']$ or contains non-climbing points$\rangle$ and $|c_l'| < c$
then make $c_l'$ helping at level $i$. If $\langle [e_2';e_2)$ is not of type
$[e_2';e_2]$ or contains non-climbing points$\rangle$ and $|c_r'| < c$ then
make $c_r'$ helping at level $i$.  Call fix$(i)$, fix$(j)$,
fix$(\min(l,i))$ and fix$(\min(i,r))$. \\

\noindent \textsl{\textsf{Guarding}} If $e = \min(P)$ or $e = \max(P)$ we
simply do nothing and return. Let $e_1 = \pred_{G_{\leq h}}(e)$ be the left
endpoint of the left interval $(e_1;e[$ lying at level $h$ and $e_2 =
\succ_{G_{\leq h}}(e)$ be the right endpoint of the right interval $[e;e_2)$
lying at level $i$, we assume w.l.o.g. that $h > i$, the case $h < i$ is
symmetric. Also let $l$ be the level of the left interval of $e_1$ and $r$ the
level of the right interval of $e_2$. Let $p_2 = \pred_{B_h \backslash G_h \cap
[e_1;e]}(p_1)$ and $p_1 = \pred_{B_h \backslash G_h \cap [e_1;e]}(e)$ be the
two left points of $e$, if they exists, $s_1 = \succ_{B_i \backslash G_i \cap
[e;e_2]}(e)$ and $s_2 = \succ_{B_i \backslash G_i \cap [e;e_2]}(s_1)$ the two
right points of $e$, if they exits.  Also let $c_l = \FGL_{C_i \cap
[e_1;e]}(e)$ and $c_r = \FGR_{C_i \cap [e;e_2]}(e)$.

If $p_2$ does not exist we make $e_1$ guarding at level $\min(l,j)$, we make
$p_1$ of type $t_{\text{after}}$ at level $j$ and let $e_1'$ denote $e_1$, else
we make $p_1$ guarding at level $j$ and let $e_1'$ denote $p_1$. If it is the
case that $i > j$ then we check: if $s_2$ does not exist then we make $s_1$ of
type $t_{\text{after}}$ at level~$j$, $e_2$ guarding at level $\min(j,r)$ and
let $e_2'$ denote $e_2$, else we make $s_1$ guarding at level $j$ and let $e_2'$
denote $s_1$. We make $e$ of type $t_{\text{after}}$ at level $j$.

Now let $c_l'$ be the points of $c_l$ which was not moved and $c_r'$ the points
of $c_r$ which was not moved. If $|c_l'| < c$ then make $c_l'$ helping at level
$h$. We now have two cases if $e_2'$ exists: then if $|c_r'| < c$ then make
$c_r'$ helping at level $i$. The other case is if $e_2'$ does not exist: then
if $\langle (e_1';e_2)$ is not of type $[e_1';e_2]$ or contains non-climbing
points$\rangle$ and $|c_r'| < c$ then make $c_r'$ helping at level $i$. In all
cases call fix$(\min(l,h))$, fix$(h)$ and fix$(i)$. If $i>j$ then call
fix$(j)$ and fix$(\min(j,r))$.

\subparagraph*{Delete$(e)$} We first call find$(e)$ to get the type of $e$ and
its level $i$, if $e$ is not in the dictionary we just return. If $e$ is in the
dictionary we have two cases, depending on if $e$ is guarding or not. \\

\noindent \textsl{\textsf{Non-guarding}} Let $c_l = \GIL_{C_i}(e)$ be the
elements in the climbing group immediately left of $e$, let $c_r =
\GIR_{C_i}(e)$ be the elements in the climbing group immediately right of $e$,
let $h_l = \GIL_{H_i}(e)$ be the elements in the helping group immediately left
of $e$, and let $h_r = \GIR_{H_i}(e)$ be the elements in the helping group
immediately right of $e$. Let $e_1 = \pred_{G_{\leq i}}(e)$ and let $e_2 =
\succ_{G_{\leq i}}(e)$. Let $l$ be the level of the interval left of $e_1$ and
$r$ the level of the interval right of $e_2$.

We have two cases, the first is $|]e_1;e_2[ \cap B_i| = 1$: if $l > r$ make
$e_1$ guarding and $e_2$ arriving at level $r$, if $l < r$ then make $e_1$
arriving and $e_2$ guarding at level $l$. If $l = r$ and $|P|=n \geq 4$ then
make $e_1$ and $e_2$ arriving at level $l=r$. Delete $e$, call fix$(r)$,
fix$(l)$, fix$(i)$ and rebalance-above$(1)$.

The other case is $|]e_1;e_2[ \cap B_i| > 1$: If $\langle(e_1;e_2)$ is not of
type $[e_1;e_2]$ or contains non-climbing points$\rangle$ and $|c_l|+|c_r| < c$
then make $c_l$ and $c_r$ helping at level $i$. If $|h_l|+|h_r| \geq c$ then
make $h_l$ and $h_r$ climbing at level $i$. Delete $e$, call fix$(i)$ and
rebalance-above$(1)$. \\

\noindent \textsl{\textsf{Min-guarding}} If $e = \min(P)$ then let $e' =
\succ_{G_{\leq m}}(e)$ and $e'' = \succ_{G_{\leq m}}(e')$ where $0$ is the
level of $(e;e')$ and $i$ is the level of $(e';e'')$. The case of $e = \max(P)$
is symmetric. Also let $s_1=\succ_{B_0\backslash G_0 \cap [e;e']}(e)$,
$s_2=\succ_{B_0\backslash G_0 \cap [e;e']}(s_1)$, $t_1=\succ_{B_i\backslash G_i
\cap [e';e'']}(e')$ and $t_2=\succ_{B_i\backslash G_i \cap [e';e'']}(t_1)$.

If $s_2$ exists then delete $e$ make $s_1$ guarding at level $0$ and call fix$(0)$. If $s_2$ does not exist and $t_2$ exists then delete $e$ make $s_1$ and $t_1$ guarding and $e'$ arriving at level $0$ and finally call fix$(0)$ and fix$(i)$. If $s_2$ does not exist and $t_2$ does not exist then delete $e$, make $s_1$ and $e''$ guarding and $e'$ and $t_1$ arriving at level $0$ and finally call fix$(0)$ and fix$(i)$. In all the previous cases return. \\

\noindent \textsl{\textsf{Guarding}} Let $h$ be the level of the left interval
$(e_1:e[$, let $i$ the level of the right interval $[e:e_2)$ that $e$
participates in. We assume w.l.o.g. that $h > i$, the case $h < i$ is
symmetric. Let $l$ the level of the left interval that $e_1$ participates in,
where $e_1 = \pred_{G_{\leq h}}(e)$ and $e_2 = \succ_{G_{\leq h}}(e)$. Let $p_2
= \pred_{B_h \backslash G_h \cap [e_1;e]}(p_1)$ and $p_1 = \pred_{B_h
\backslash G_h \cap [e_1;e]}(e)$.  Let $c_l = \FGL_{C_i}(e)$ be the points in
the first group of climbing points left of $e$.

If $p_2$ exist we make $p_1$ guarding at level $i$, and let $e'$ denote $p_1$,
else we make $e_1$ guarding at level $\min(l,i)$, let $e'$ denote $e_1$ and if
$[e';e_2)$ is of type $[e';e_2]$ and contains only climbing points then we make
$p_1$ climbing at level $i$ else we make $p_1$ waiting at level $i$. Let $c_l'$
be the points in $c_l$ which was not moved in the previous movement of points.
If $|c_l'| < c$ make $c_l'$ helping at level $h$. If $e'$ is $e_1$ then call
fix$(l)$. Delete $e$, call fix$(h)$, fix$(i)$ and rebalance-above$(1)$.

\subparagraph*{Rebalance-below$(i)$} For each level $l=0,\ldots,i$ we perform a
shift-up$(l)$ while $c < c_l$.

\subparagraph*{Rebalance-above$(i)$}  For each level $l=i,\ldots,m-1$ we
perform shift-down$(l+1)$ while $c_l < -c$.

%%%%%%%%%%%%%%%%%%%%%%%%%%%%%%%%%%%%%%%%%%%%%%%%%%%%%%%%%%%%%%%%%%%%%%%%%%%%%%%%
\section{Memory management} \label{sec:memoryManagement}
%%%%%%%%%%%%%%%%%%%%%%%%%%%%%%%%%%%%%%%%%%%%%%%%%%%%%%%%%%%%%%%%%%%%%%%%%%%%%%%%

We will now deal with the memory layout of the data structure. We will put the
blocks in the order $B_0,\ldots,B_m$, where block $B_i$ further has its
dictionaries in the order $D_i, A_i, R_i, W_i, H_i, C_i$ and $G_i$, see Figure
\ref{fig:memoryOverview}. Block~$B_m$ grows and shrinks to the right when
elements are inserted and deleted from the working set dictionary.

The $D_i$ structure is not a moveable dictionary as the other structures in a
block are, it is simply an array of $w_i = d 2^{i+k}$ elements which we use to
encode the size of each of the structures $A_i, R_i, W_i, H_i, C_i$ and $G_i$
along with their own auxiliary data, as they are not implicit and need to
remember $\bigO(2^{i+k})$ bits which we store here. As each of the moveable
dictionaries in $B_i$ have size $\bigO(2^{2^{i+k}})$ we need to encode numbers
of $\bigO(2^{i+k})$ bits in $D_i$.

We now describe the memory management concerning the movement, insertion and
deletion of elements from the working-set dictionary. First notice that the
methods find, predecessor and successor do not change the working-set
dictionary, and layout in memory. Also the methods shift-down, search,
rebalance-below and rebalance-above only calls other methods, hence their memory
management is handled by the methods they call. The only methods where actual
memory management comes into play are in insert, shift-up, fix, move-down and
delete. We will now describe two methods internal-movement -- which handles
movement inside a single block/level -- and external-movement -- which handles
movement across different blocks/levels. Together these two methods handle all
memory management.

\subparagraph*{Internal-movement$(m_1,\ldots,m_l)$} Internal-movement in level
$i$ takes a list of \emph{internal moves} $m_1,\ldots,m_l$ to be performed on
block $B_i$, where $l = \bigO(1)$ and move $m_j$ consists of:
\begin{itemize}
  \item the index $\gamma = D_i, A_i, R_i, W_i, H_i, C_i, G_i$ of the dictionary
  to change, where we assume\footnote{We will misuse notation and let $\gamma+1$
  denote the next in the total order $D, A, R, W, H, C, G$. We will also compare
  $m_j.\gamma$ and $m_h.\gamma$ with $\leq$ in this order.} that $m_j.\gamma
  \leq m_h.\gamma$, for $j \leq h$,

  \item the set of elements $S_\text{in}$ to put into $\gamma$, where
  $|S_\text{in}| = \bigO(1)$,

  \item the set of elements $S_\text{out}$ to take out of $\gamma$, where
  $|S_\text{out}| = \bigO(1)$ and

  \item the total size difference $\delta = |S_\text{in}| - |S_\text{out}|$ of
  $\gamma$ after the move.
\end{itemize}
For $j=1,\ldots,l$ do: if $m_j.\delta < 0$ then remove $S_\text{out}$
from $\gamma$, insert $S_\text{in}$ into $\gamma$ and move $\gamma+1,\ldots, G$
left $|m_j.\delta|$ positions, where we move them in the order $\gamma+1,\ldots,
G$. If $m_j.\delta > 0$ then move $\gamma+1,\ldots,G$ right $m_j.\delta$
positions, where we move them in the order $G,\ldots, \gamma+1$, remove
$S_\text{out}$ from $\gamma$ and insert $S_\text{in}$ into $\gamma$. See Figure
\ref{fig:in-ex-movement}.

It takes $\bigO(\log(2^{2^{i+k}})) = \bigO(2^{i+k})$ time and
$\smash{\bigO(\log_B(2^{2^{i+k}})) = \bigO(\frac{2^{i+k}}{\log B})}$
cache-misses to perform move $j$. In total all the moves $m_1,\ldots,m_l$ use
$\bigO(2^{i+k})$ time and $\bigO(\frac{2^{i+k}}{\log B})$ cache-misses, as $l =
\bigO(1)$.

\fig{tb}{}{width=\linewidth}{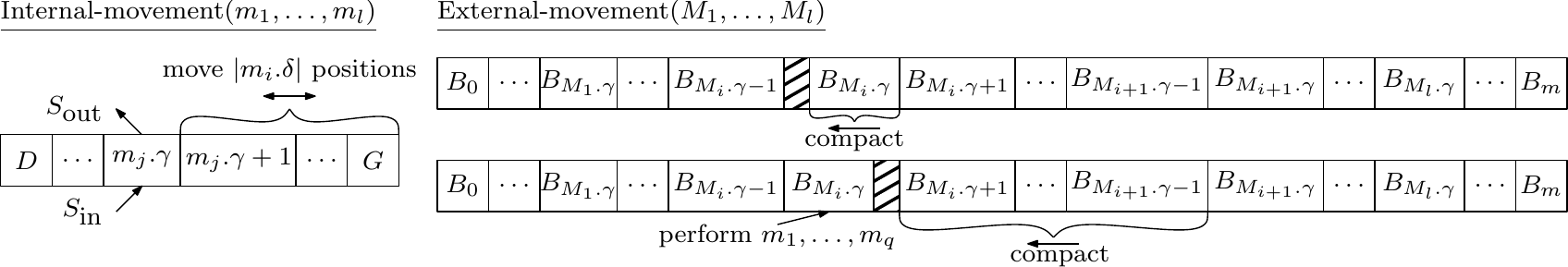}{fig:in-ex-movement}{(Left)
Memory movement of internal-movement inside of a block $B_i$. (Right) Memory
movement of external-movement across multiple blocks
$B_{M_1.\gamma},\ldots,B_{M_l.\gamma}$.}

\subparagraph*{External-movement$(M_1,\ldots,M_l)$} External-movement takes a
list of \emph{external moves}~$M_1, \ldots,$ $M_l$, where $l = \bigO(1)$. Move
$M_j$ consists of:
\begin{itemize}
  \item the index $0\leq \gamma \leq m$ of the block/level to perform the
  internal moves $m_1,\ldots,m_q$ on, where $M_j.\gamma < M_h.\gamma$ for $j
  < h$,

  \item the list of internal moves $m_1,\ldots,m_q$ to perform on block
  $\gamma$, where $q = \bigO(1)$, and

  \item the total size difference $\Delta = \sum_{h=1}^q{m_h.\delta}$ of block
  $\gamma$ after all the internal moves $m_1,\ldots,m_q$ have been performed.
\end{itemize}
Let $\overline{\Delta} = \sum_{i=1}^l{M_i.\Delta}$ be the total size change of
the dictionary after the external-moves have been performed. If
$\overline{\Delta} = 0$ then we let $\gamma_\text{end} = M_l.\gamma$ else we
let $\gamma_\text{end} = m$. Let~$p_\text{end} =
\sum_{j=0}^{\gamma_\text{end}}{|B_j|}+\overline{\Delta}$ be the last address of
the right most block that we need to alter. Let $s_1,\ldots,s_k$ be the sublist
of the indexes $\{1,\ldots,l\}$ where $M_{s_i}.\Delta \leq 0$ for
$i=1,\ldots,k$.  Let $a_1,\ldots,a_h$ be the sublist of the indexes
$\{1,\ldots,l\}$ where $M_{a_i}.\Delta > 0$ for $i=1,\ldots,h$.

We first perform all the internal moves of each of the external moves
$M_{s_1},\ldots,M_{s_k}$. Then we compact all the blocks with index $i$ where
$M_1.\gamma \leq i \leq \gamma_\text{end}$ so the rightmost block ends at
position $p_\text{end}$.  Finally for each external move $M_{a_i}$ for
$i=1,\ldots,h$: move $B_{M_{a_i}.\gamma}$ left so it aligns with
$B_{M_{a_i}.\gamma-1}$ and perform all the internal moves of $M_{a_i}$, then
compact the blocks $B_{M_{a_i}.\gamma+1}, \ldots, B_{M_{a_{i+1}}.\gamma-1}$ at
the left end so they align with block $B_{M_{a_i}.\gamma}$.

It takes $\bigO\(l\log\(2^{2^{i+k}}\)\) = \bigO\(l2^{i+k}\)$ time and
$\bigO\(l\log_B\(2^{2^{i+k}}\)\) = \bigO\(l\frac{2^{i+k}}{\log B}\)$
cache-misses to perform the internal moves on level $i$. In total all the
external moves $M_1,\ldots,M_l$ use $\bigO(2^{\gamma_\text{end}+k})$ time and
$\bigO\(\frac{2^{\gamma_\text{end}+k}}{\log B}\)$ cache-misses, as the external
move at level $\gamma_\text{end}$ dominates the rest and $l = \bigO(1)$.

\subsection{Memory management in updates of intervals}

With the above two methods we can perform the memory management when updating
the intervals in Section \ref{sec:operations}: Whenever an element moves
around, is deleted or inserted, it is simply put in one or two internal moves.
All internal moves in a single block/level are grouped into one external move.
Since all updates of intervals only move around a constant number of elements,
the requirements for internal/external-movement that $l = \bigO(1)$ and $q =
\bigO(1)$ are fulfilled. From the above time and cache bounds for the memory
management the bounds in Theorem \ref{thm:results} follows.

%%%%%%%%%%%%%%%%%%%%%%%%%%%%%%%%%%%%%%%%%%%%%%%%%%%%%%%%%%%%%%%%%%%%%%%%%%%%%%%%
\begin{fullenv}
\section{Analysis} \label{sec:analysis}
%%%%%%%%%%%%%%%%%%%%%%%%%%%%%%%%%%%%%%%%%%%%%%%%%%%%%%%%%%%%%%%%%%%%%%%%%%%%%%%%

We will leave it for the reader to check that the pre-conditions for each
methods in Section \ref{sec:operations} are fulfilled and that the methods
maintains all invariants. We will instead concentrate on using the invariants
to prove correctness of the find, predecessor, successor and shift-up
operations along with proving time and cache-miss bounds for these. We will
leave the time and cache-miss bounds of search, rebalance-above,
rebalance-below, shift-down, insert, delete and fix for the reader as they are
all similarly in nature.

\subparagraph*{Find$(e)$} We only consider the cases where $\min(P) < e <
\max(P)$, the other cases trivially gives the correct answer in $\bigO(1)$ time
and cache-misses as $\min(P), \max(P) \in G_0$. Assume that find$(e)$ stops at
level $i$, then we have that $e_1 \leq p$ or $s \leq e_2$ so $I(e_1,e_2,i) \neq
\emptyset$ and $i$ is the minimal $i$ where this happens, see lemma
\ref{lem:intervals}. Notice that $e_1 = \pred_{G_{\leq i}}(e)$ and $e_2 =
\succ_{G_{\leq i}}(e)$, so $e_1$ and $e_2$ are the same as in lemma
\ref{lem:intervals}. When the while loop breaks we have all the
preconditions for lemma \ref{lem:intervals}. Now $e$ is either in the
dictionary, or not, and if $e$ is in the dictionary it is either guarding or
not, so we have three cases.

Case 1) $e$ is in the dictionary and is non-guarding: then we have from lemme
\ref{lem:intervals} that $(e_1;e_2)$ is a interval at level $i$ and $e \in B_i$.
From this we also have that $\log(\ws{e}) \geq \log(2^{2^{i+k-1}})$.

Case 2) $e$ is not in the dictionary: from lemma \ref{lem:intervals} $(e_1;e_2)$
lie at level $i$ and we know that $e$ intersects it. Since $e$ is not in the
dictionary $\ws{e} = n$ and then $\log(\ws{e}) \geq \log(2^{2^{i+k-1}})$.

Case 3) $e$ is in the dictionary and is guarding: from lemma \ref{lem:intervals}
we have that either $(e_1;e)$ or $(e;e_2)$ lie in level $i$, hence $e \in G_i
\subseteq B_i$. From this we also have that $\log(\ws{e}) \geq
\log(2^{2^{\max(i,j)+k-1}}) \geq \log(2^{2^{i+k-1}})$.

From the above we see that find$(e)$ runs in $\bigO(\log(2^{2^{i+k-1}})) = \bigO(\log
\min(\ws{\pred(e)},\ws{e}, \ws{\succ(e)}))$ time. When we look at the
cache-misses we will first notice that the first $\floor{\log \log B}$ levels
will fit in a single cache-line because all levels are next to each other in the
memory layout, so the total cache-misses will be
\[
  \bigO\(1+\sum_{j=\floor{\log \log B}+1}^i{\(1+\log_B\(2^{2^{j+k}}\)\)}\) =
  \bigO\(\frac{2^{i+k}}{\log B}\) = \bigO(\log_B \min(\ws{\pred(e)},\ws{e},
  \ws{\succ(e)})).
\]

\subparagraph*{Predecessor$(e)$ (and successor$(e)$)} We will only handle the
predecessor operation, the case for the successor is symmetric. Since we have
the same condition in the while loop as for find, we know that when it breaks
it implies that $I(e_1,e_2,i) \neq \emptyset$. So from lemma
\ref{lem:intervals}, $e$ intersects a interval at level $i$ and the predecessor
of $e$ is now $\max(e_1,p)$.

From \iref{it:wsvalue} we know that $\log(\ws{p}) \geq \log(2^{2^{i+k-1}})$ and
the total time usage is $\sum_{j=0}^i{\bigO(\log(2^{2^{i+k}}))}$ $=
\bigO(2^{i+k}) = \bigO(\log(\ws{p}))$. Like in find, the first $\floor{\log
\log B}$ levels fit into one block/cache-line hence the total cache-misses will
be $\bigO(\log_B(\ws{p}))$.

\subparagraph*{Shift-up$(i)$} For shift-up to work for level $i$ it is
mandatory that $|C_i| > 0$ so that $\succ_{C_i}(-\infty)$ will return a element
which can be moved to level $i+1$. From the precondition that $|H_i| + |C_i| =
4c2^{2^{i+k}} + c'_i$, where $c \leq c'_i = \bigO(1)$, we have that
\[
  |C_i| = 4c2^{2^{i+k}} + c'_i - |H_i| \geq 4c2^{2^{i+k}} - c - |H_i|
\]
so proving that $|H_i| < 4c2^{2^{i+k}} - c$ is enough. From \iref{it:hneighbors}
we can at most have $c-1$ helping points in a helping group, so for every $c-1$
helping points we need a separating point, the role of the separating point can
be played by a point from $D_i,A_i,R_i,W_i$ or $G_{\leq i-1}$. These are the
only ways to contribute points to $H_i$ hence for $i\geq 1$ we have this bound
\begin{eqnarray*}
  |H_i| &\leq& (c-1)(|D_i|+|A_i|+|R_i|+|W_i|+|G_{\leq i-1}|) \\
  & \stackrel{(*)}{\leq} &
  (c-1)\(w_i + 2\cdot 2^{2^{i+k}} +
  \sum_{j=0}^{i-1}{\((4+2d+8c)2^{2^{j+k}} + 2c\)}\) \\
  & \stackrel{(**)}{\leq} &
  (c-1)\(d \cdot 2^{i+k} + 2\cdot 2^{2^{i+k}} + (4+2d+8c)\cdot 2 \cdot
  2^{2^{i+k-1}} + 2ci\)
\end{eqnarray*}
Where we in $(*)$ have used \iref{it:encode}, \iref{it:resting}
\iref{it:arrivingwaiting} and \oref{it:limitedguarding}, and in $(**)$ have
used that $2^{2^l} = 2^{2^{l-1}} \cdot 2^{2^{l-1}}$ and $2^{2^{l-1}} \geq l$
for $l\geq 1$. If we use that $c=5$ then for $k > \log \log (380+20d) + 1$ we
have that $|C_i| \geq 4c2^{2^{i+k}} - c - |H_i| > 0$ for $i=1,\ldots,m-1$.

For $i=0$ we have a different bound as $G_{\leq i-1}$ is empty, we get the bound
\begin{eqnarray*}
  |H_0| &\leq& (c-1)(|D_i|+|A_i|+|R_i|+|W_i|) \\
  & \leq &
  (c-1)\(d \cdot 2^{i+k} + 2\cdot 2^{2^{i+k}}\)
\end{eqnarray*}
but for $k > \log \log (380+20d) + 1$ this is of course still sufficient as
$|H_0|$ only got smaller. So we have proved that $|C_i| > 0$ for level
$i=0,\ldots,m-1$.

\subparagraph*{Move-down$(e,i,j,t_{\text{before}},t_{\text{after}})$} Move-down
moves a constant number of points around and into level $j$ from $i$.  If $e$
is non-guarding we call fix$(i)$, fix$(j)$, fix$(\min(l,i))$ and
fix$(\min(i,r))$. If $e$ is guarding we call fix$(\min(l,h))$,
fix$(h)$ and fix$(i)$, and if $i>j$ we also call fix$(j)$ and
fix$(\min(j,r))$.  In the non-guarding case the time is bounded by
$\bigO(\log 2^{2^{i+k}}) = \bigO(\log \ws{e})$ and the cache-miss bounds are
dominated by $\bigO(\log_B 2^{2^{i+k}}) = \bigO(\log_B \ws{e})$. In the
guarding case the time is bounded by $\bigO(\log 2^{2^{h+k}}) = \bigO(\log
\ws{e})$ and the cache-miss bounds are dominated by $\bigO(\log_B 2^{2^{h+k}})
= \bigO(\log_B \ws{e})$.

%%%%%%%%%%%%%%%%%%%%%%%%%%%%%%%%%%%%%%%%%%%%%%%%%%%%%%%%%%%%%%%%%%%%%%%%%%%%%%%%
\end{fullenv}
%%%%%%%%%%%%%%%%%%%%%%%%%%%%%%%%%%%%%%%%%%%%%%%%%%%%%%%%%%%%%%%%%%%%%%%%%%%%%%%%
\begin{fullenv}
\section{Further work}
%%%%%%%%%%%%%%%%%%%%%%%%%%%%%%%%%%%%%%%%%%%%%%%%%%%%%%%%%%%%%%%%%%%%%%%%%%%%%%%%

We still have some open problems. Is it possible to change the insert operation
such that when we insert a new point it will get a working-set value of $n+1$
instead of $0$? We can actually achieve this in our structure by loosening the
invariant on the working-set number of guarding points to only require that they
have a working-set number of at least $2^{2^{\min(i,j)+k-1}}$, but then for
search the time will increase to $\bigO(\log \min(\ws{e},
\max(\ws{\pred(e)},\ws{\succ(e)})))$ and the cache-misses to $\bigO(\log_B
\min(\ws{e}, \max(\ws{\pred(e)},\ws{\succ(e)})))$ and the bounds for predecessor
and successor queries would increase to $\bigO(\log
\max(\ws{\pred(e)},\ws{\succ(e)}))$ time and $\bigO(\log_B
\max(\ws{\pred(e)},\ws{\succ(e)}))$ cache-misses.

Another interesting question is if we can have a dynamic dictionary supporting
efficient finger searches \cite{B05} in the implicit model, i.e., we have a
finger $f$ located at a element and then we want to find an element $e$ in time
$\bigO(\log d(f,e))$, where $d(f,e)$ is the rank distance between $f$ and $e$.
But very recently \cite{NT11} have shown that finger search in $\bigO(\log
d(e,f))$ time is not possible in the implicit model. They give a lower bound of
$\Omega(\log n)$. Now we could instead separate the finger search and the
update of the finger, say we allow the finger search to use $\bigO(q(d(e,f)))$
time for some function $q$. In this setting they also prove a lower of
$\Omega(q^{-1}(\log n))$ for the update finger operation, where $q^{-1}$ is the
inverse function of $q$. They also give almost tight upper bounds for this
setting, in the form of a trade-off bound between the finger search and the
update finger operations. The finger search operation uses $\bigO(\log d(e,f))
+ q(d(e,f))$ time, and the update finger operation uses $\bigO(q^{-1}(\log n)
\log n)$ time. But even given their result it still remains an open problem
whatever dynamic finger search with an externally maintained finger is possible
in $\bigO(\log d(e,f))$ time. So in other words is it possible to do finger
search in $\bigO(\log d(e,f))$ time if we allow the data structure to store
$\bigO(\log n)$ bits of data that can store the finger?

%%%%%%%%%%%%%%%%%%%%%%%%%%%%%%%%%%%%%%%%%%%%%%%%%%%%%%%%%%%%%%%%%%%%%%%%%%%%%%%%
\end{fullenv}
%%%%%%%%%%%%%%%%%%%%%%%%%%%%%%%%%%%%%%%%%%%%%%%%%%%%%%%%%%%%%%%%%%%%%%%%%%%%%%%%
% Litteraturliste
\fullcmt{\clearpage}
\bibliography{Article}
%%%%%%%%%%%%%%%%%%%%%%%%%%%%%%%%%%%%%%%%%%%%%%%%%%%%%%%%%%%%%%%%%%%%%%%%%%%%%%%%

%%%%%%%%%%%%%%%%%%%%%%%%%%%%%%%%%%%%%%%%%%%%%%%%%%%%%%%%%%%%%%%%%%%%%%%%%%%%%%%%
\end{document}